\documentclass[sigconf]{acmart}

\usepackage{natbib}
\usepackage{url}
\usepackage{balance}
\usepackage{subcaption}
\usepackage[font=small]{caption}
\usepackage[inline]{enumitem}

\newcommand{\m}[1]{#1\:\mathrm{m}}
\newcommand{\kg}[1]{#1\:\mathrm{kg}}
\newcommand{\V}[1]{#1\:\mathrm{V}}
\newcommand{\A}[1]{#1\:\mathrm{A}}
\newcommand{\GHz}[1]{#1\:\mathrm{GHz}}
\newcommand{\MHz}[1]{#1\:\mathrm{MHz}}
\newcommand{\W}[1]{#1\:\mathrm{W}}
\newcommand{\kgf}[1]{#1\:\mathrm{kgf}}
\newcommand{\us}[1]{#1\:\mathrm{\mu s}}

\title{\textit{What is A Wireless UAV?}\\ A Design Blueprint for 6G Flying Wireless Nodes}

\author{John Buczek,
Lorenzo Bertizzolo,
Stefano Basagni,
Tommaso Melodia}
\affiliation{
 \institution{Institute for the Wireless Internet of Things, Northeastern University, Boston, MA 02115, USA }
Email: \{buczek.j, bertizzolo.l, s.basagni, melodia\}@northeastern.edu}
\thanks{This article is based upon material supported in part by the US National Science Foundation under grant CNS \#1923789.}


\begin{abstract}
Wireless Unmanned Aerial Vehicles (UAVs) were introduced in the world of $4$th generation networks (4G) as cellular users, and have attracted the interest of the wireless community ever since. 
In~5G, UAVs operate also as flying Base Stations providing service to ground users.
They can also implement independent off-the-grid UAV networks. 
In~6G networks, wireless UAVs will connect ground users to in-orbit wireless infrastructure.
As the design and prototyping of wireless UAVs are on the rise, the time is ripe for introducing a more precise definition of what is a wireless UAV.
In doing so, we revise the major design challenges in the prototyping of wireless UAVs for future 6G spectrum research. 
We then introduce a new wireless UAV prototype that addresses these challenges. %
The design of our wireless UAV prototype will be made public and freely available to other researchers. 
\end{abstract}


\begin{document}

\acmYear{2022}\copyrightyear{2022}
\setcopyright{acmlicensed}
\acmConference[WiNTECH '21]{The 15th ACM Workshop on Wireless Network Testbeds, Experimental evaluation \& CHaracterization}{January 31--February 4, 2022}{New Orleans, LA, USA}
\acmBooktitle{The 15th ACM Workshop on Wireless Network Testbeds, Experimental evaluation \& CHaracterization (WiNTECH '21), January 31--February 4, 2022, New Orleans, LA, USA}
\acmPrice{15.00}
\acmDOI{10.1145/3477086.3480840}
\acmISBN{978-1-4503-8703-3/22/01}

\maketitle

\section{Introduction}


Unmanned Aerial Vehicles (UAVs, also known as \emph{drones}) have progressively attracted the interest of the wireless community as a tool to deploy flexible and on-demand network infrastructure.
During the last decade, several bodies and stakeholders have worked to regulate and facilitate the use of UAVs with the goal of promoting their integration into  current and next generation networks. 
Among others, these include networking regulatory bodies (e.g., the 3GGP and the FCC), cellular wireless carriers (e.g., AT\&T, China Mobile, and Vodafone), aviation authorities (e.g., the FAA), and large corporations ready to employ flying connected robots for their business operations (e.g., Amazon, DHL and CNN)~\cite{bertizzolo2021streaming}.
%

The main applications of wireless UAVs as networking nodes include the following:

\begin{itemize}[leftmargin=*]

\item \textit{UAVs for networking and service provisioning}: UAVs implement flying Base Stations (BSs) that connect users on the ground to the Internet. 
UAVs here extend missing or malfunctioning cellular infrastructure by routing the traffic of unserved ground cellular users (User Equipments, or UEs) to an Internet gateway, like the nearest cell tower.
As such, the UAVs implement both wireless access and backhaul infrastructure~\cite{sheshadri2020skyhaul}. 

\item \textit{UAVs as flying cellular users}: UAVs are flying robots connected to the ground cellular network. They can be controlled and commanded over the Internet, and do not require a flight operator in the vicinity. Additionally, they can upload drone-sourced sensor data and video streaming to a cloud server using cellular connectivity~\cite{bertizzolo2021streaming}.

\item \textit{UAV swarms for on-demand, private service networks}: Multiple wireless UAVs will form a flying mesh network to connect users on the ground with each other. This way, UAVs implement a private service network that can be deployed on-demand as an alternative to the unavailable or untrustworthy cellular infrastructure~\cite{BertizzoloInfocom20SwarmControl, cheng2021learning}.

\item \textit{UAVs for 6G}: UAVs will implement key 6G network infrastructure operating in the low atmosphere ($<1\mathrm{\:km}$) which will connect ground users with the non-terrestrial 6G infrastructure deployed in-orbit ($20$ to~$35786\mathrm{\:km}$)~\cite{geraci2021will}.

\end{itemize}

\noindent
Due to the impracticality of employing power lines or fiber links on a flying node for any of the above applications, wireless UAVs differ from most of the network infrastructure: they are fully wireless---in control, radio access, and backhaul. 
Additionally, UAVs are exclusively battery-powered. 
Thus, the design of wireless UAV testing platforms for future wireless research requires a deep understanding of interactions among the powering, wireless and motion capabilities of UAVs.
While recent years witnessed a surge in wireless UAVs-related research, works that test their solutions on real UAV hardware are limited to a handful. 
Authors mainly employed UAV hardware available on the market and equipped it with compute and wireless capabilities to satisfy their needs, leaving many designs and prototyping questions unanswered.

\textit{We believe the time is ripe to provide further insights on what a wireless UAV is and to explain in detail what are the key design choices to be made toward the prototyping of future wireless UAV testing platforms for~6G-related research.}

Our work provides the first formal definition of wireless UAV.
We go over key design principles for the design and development of future wireless UAV testing platforms, and we introduce a new wireless UAV prototype that meets the requirements of those principles.
The wireless UAV that we present in this article has been carefully designed to address the present and future challenges for 6G spectrum research. 
Its design process, components and assembling instructions are made public for the community to use.


\section{Wireless UAVs: A Formal Definition}

Here and for the first time, we formally enunciate the design blueprint of a Wireless UAV.
While wireless UAVs can be employed for different wireless applications, vary in size, or in manufacturer, they should include the following.

\begin{enumerate}[leftmargin=*]

    \item \textit{Frame}: This is the physical structure of the robot, and concerns the main plane, the motors, the electronic speed controllers (ESC), the propellers, the batteries, and all the other structural components of a flying robot. 
    
    \item \textit{Computing unit}: Off-the shelf or added on, future wireless UAVs should be equipped with a computing unit comprehensive of CPU, memory, RAM, disk, GPU, hardware and software architecture to run an Operative System. 
    
    \item \textit{Flight Control Unit (FCU)}: The FCU controls the mobility of the UAV by driving its individual motors. The FCU hosts a flight controller firmware (FCF) that regulates the electricity fed into the motors by the ESCs.
    
    \item \textit{Radio front-ends}: To implement wireless communication, wireless UAVs feature one or more radio front-ends on board. This can include Wi-Fi or Bluetooth chip-sets, cellular modems or programmable Software-defined Radios (SDR). These modules include the antennas and the on-board powering circuitry. 
    
    \item \textit{Wireless stack}: If the radio front-ends are in charge of modulating and demodulating electromagnetic signals into information bits, upper-layers operations are handled by the wireless stack. 
    This can be partially implemented in hardware or all in software. 
    The wireless stack is in charge of implementing Physical and MAC layer framing and functionalities, handling traffic forwarding, operating session management and implement all the protocol stack functionalities to supports the application layer.
    The wireless stack is implemented on the on-board computing unit for its software implementation side and on the radio front ends for its hardware implementation side. 
    When employing SDR as a radio front end, the implementation if entirely in software. 
    
    \item \textit{Control APIs}: Future wireless UAVs will have to expose control of their wireless and motion functionalities in (near) real time. Accordingly, the radio front-end(s), the wireless stack, and the FCU must expose \textit{control APIs}.
    Through the control APIs, a control program running on the on-board computer, or communicating with it, can change the wireless UAV operational parameters for a specific control objective in (near) real time.
    
\end{enumerate}

According to this definition of a wireless UAV we classify the most important related work on the design and prototyping of wireless UAV testing platforms. 
Table~\ref{table:relatedwork} reports wireless UAV prototypes indicating frame, FCU, computing unit and radio front-end.

\begin{table*}[t]
\centering
\small
\begin{tabular}{ |c||c|c|c|c| } 
\hline
Paper & UAV model & FCU & Computing unit & Radio front-end \\
\hline
\hline
\cite{BertizzoloInfocom20SwarmControl} & Intel Aero & Pixhawk for Intel Aero / PX4 & Intel Aero Board & USRP B205-mini-i\\
\cite{bertizzolo2021streaming} & DJI M100 &  Pixhawk / PX4 & Intel NUC & USRP B210 \\
\cite{polese2020experimental, BertizzoloMmnets19, sanchez2020robust, sanchez2020millimeter} & DJI M600 Pro & DJI FCU & Intel NUC & Facebook Terragraph  \\
\cite{BertizzoloHotmobile20} & Intel Aero &  Pixhawk for Intel Aero / PX4 & Intel Aero Board & ZTE LTE USB dongle  \\
\cite{sheshadri2020skyhaul} & DJI M600 & DJI FCU & \textit{Unknown} (Intel Core i7-6600U CPU) & MikroTik WAP 60G  \\
\cite{moradi2018skycore} & DJI M600 & DJI FCU & \textit{Unknown} (4-core, 1.9 GHz CPU, 8 GB RAM) & S60 nanoLTE  \\
\cite{ferranti2020skycell} & DJI M600 & DJI FCU & Intel NUC & USRP B210  \\
\cite{amorim2017radio} & DJI M600 & DJI FCU & --- & R\&S TSMA   \\
\cite{chakraborty2018skyran} & DJI M600 & DJI FCU & \textit{Unknown} (Intel Core i7) + \textit{Unknown} (J1900 CPU) & USRP B210\\
\cite{amorim2018measured} & DJI M600 & DJI FCU & --- &R\&S QualiPoc Smartphone  \\
\cite{d2019quality} & DJI M100 & Pixhawk / PX4 & Intel NUC & USRP B200mini-i   \\
\cite{ferranti2019hiro} &  Intel Aero & Pixhawk for Intel Aero / PX4 & Intel Aero Board &  NXP i.MX7D  \\
\cite{laclau2021signal} & Crazyflie  2.1  &  Crazyflie Bolt & Crazyflie Board & Crazyradio  \\
\cite{mishra2020survey} &  DJI Phantom 3 & DJI FCU & Raspberry Pi 3B  & --- / VNF  \\
\cite{khawaja2017uav} & DJI S-1000  & DJI FCU & --- & PEM009-KIT \& USRP X310  \\
\cite{sundqvist2015cellular} & Custom & Pixhawk / PX4 & Raspberry Pi 3B & LTE USB dongle  \\
\cite{muzaffar2020first} & Asctec Pelican & Asctec Atomboard & Asctec Board & Wistron NeWeb mobile platform  \\
\cite{burke2019safe} & Custom & Omnibus F4 Pro & Raspberry Pi Zero & Verizon USB dongle 730L  \\
\cite{petrolo2018astro} & Custom & Pixhawk / PX4 & Zynq 7030 & Iris-030 SDR  \\
\cite{semkin2021lightweight} &  DJI M210  & DJI FCU & Intel compute stick & Spectrum Master MS2760A-0100   \\
\cite{mohanti2019airbeam} &  DJI M100  & DJI FCU & ARMv8 NVIDIA TX2 & USRP B210   \\
\hline
\textbf{This work} &  Custom  & Pixhawk / Ardupilot & Intel NUC & USRP B210   \\
\hline
\end{tabular}
\caption{Literature review of wireless UAV models, their FCU, computing, and radio front-end components.}
\label{table:relatedwork}
\end{table*}

\section{A Wireless UAV Prototype for Future 6G Wireless Research}

\subsection{Key design principles}

Given the definition of wireless UAV, in this section we outline key design principles for prototyping future wireless UAVs for 6G spectrum research.

\begin{enumerate}[leftmargin=*,label=\roman*), wide]

    \item \textit{Software-defined motion control}:
    The FCU must expose a wide range of motion control APIs. These APIs should be made accessible via a standardized interface (e.g., UART or USB), and should be controllable by software running on the on-board computer.
    The flight control firmware should be open-source to allows for modification of flight control's primitives, if needed, and guarantee operators' privacy. Additionally, it should expose on-board sensors' readings to the on-board computing unit in real time.
    
    \item \textit{Software-defined RF front-end}:
    Future 6G non-terrestrial nodes will be extremely constrained regarding the payload size and weight. Additionally, hardware replacements and new hardware rollout will be reduced to the minimum. Future non-terrestrial networks are thus being designed with programmable hardware in mind, and Software-defined Radios (SDR) are the designated radio boards to implement wireless communications. 
    Different from hardware implementations `baked' into the chipsets, SDR allow for full protocol implementation programmability, in software. Additionally, SDR allow for fine-tuning of a wide range of wireless parameters in real time. 
    As wireless protocol will have to be re-designed and adjusted to the specific non-terrestrial deployment scenario, employing SDR is paramount for the success of future 6G networks.
    
    \item \textit{Multi-connectivity}: Alongside in-orbit satellites and cell towers on the ground, future wireless UAVs will be integral in multi-layered hierarchical 6G networks \cite{wang2020potential}. To relay traffic from the ground to the orbit infrastructure and viceversa, the availability of multi-connectivity on board will be paramount. Multi-connectivity can be implemented, for example, by multiple TX and RX chains on the same radio front end, or by multiple radio front-ends on board. Multiple TX and RX chains can additionally be employed to implement robust MIMO communications, which will be needed to cover the long distances of aerial connectivity.
    
    \item \textit{Software-programmable wireless stack}:
    Similar to the previous point, the protocol implementations of the upper layers of the wireless stack must be re-programmable. MAC, Networking, Transport, and Session-layer functionalities must allow for swift software updates, new protocols roll out, and tuning of their parameters in real time, in software. 
    
    \item \textit{Control plane on board}:
    A key aspect of future 6G networks is the presence of intelligence at the edge of the network. In a nutshell, this means that the control and optimization of the operations of wireless nodes---formerly delegated to a central controller in the cloud---will be carried out at the very edge nodes of the network: the network nodes. In SDN terms, this concept translates into having a control plane on board of UAVs. 
    The control plane consists in a set of algorithms that control and optimize the data handling operations of the wireless stack and the motion operations of the flight controller.
    
    \item \textit{On-board computing unit}: 
    Intelligence at the edge of the network goes along with edge computing. Accordingly, the UAVs' computing unit must feature high-bandwidth hardware and software necessary to support a wide range of control capabilities.
    The computing unit must be powerful enough to support the previous points: drive the FCU and the radio front end(s), implement fully programmable wireless protocol stacks, and execute intelligent control logic on board. Additionally, it should feature a powerful enough GPU to support the latest cutting edge AI algorithms.
    
    \item \textit{Mobile powering}:
    Even though tethered options have been proposed in literature, a fully scalable wireless UAV fabric should operate with on-board-powered hardware only.
    This should include powering of the compute unit and the radio front end(s) without posing excessive constraint on the flight time.

    \item \textit{Electromagnetic noise}:
    The proximity of radio front ends, computing unit, batteries, power amplifiers, and sensors, might result in undesirable operational points with effects from noisy wireless communications to impaired sensor readings. Consequences can be as harsh as self-jamming to unstable flight operations. 
    In designing a wireless UAV, it is important to keep in mind the electromagnetic effects that might impact the different hardware components on board.  
\end{enumerate}

\begin{figure}[ht]
    \centering
    \includegraphics[width=0.85\columnwidth]{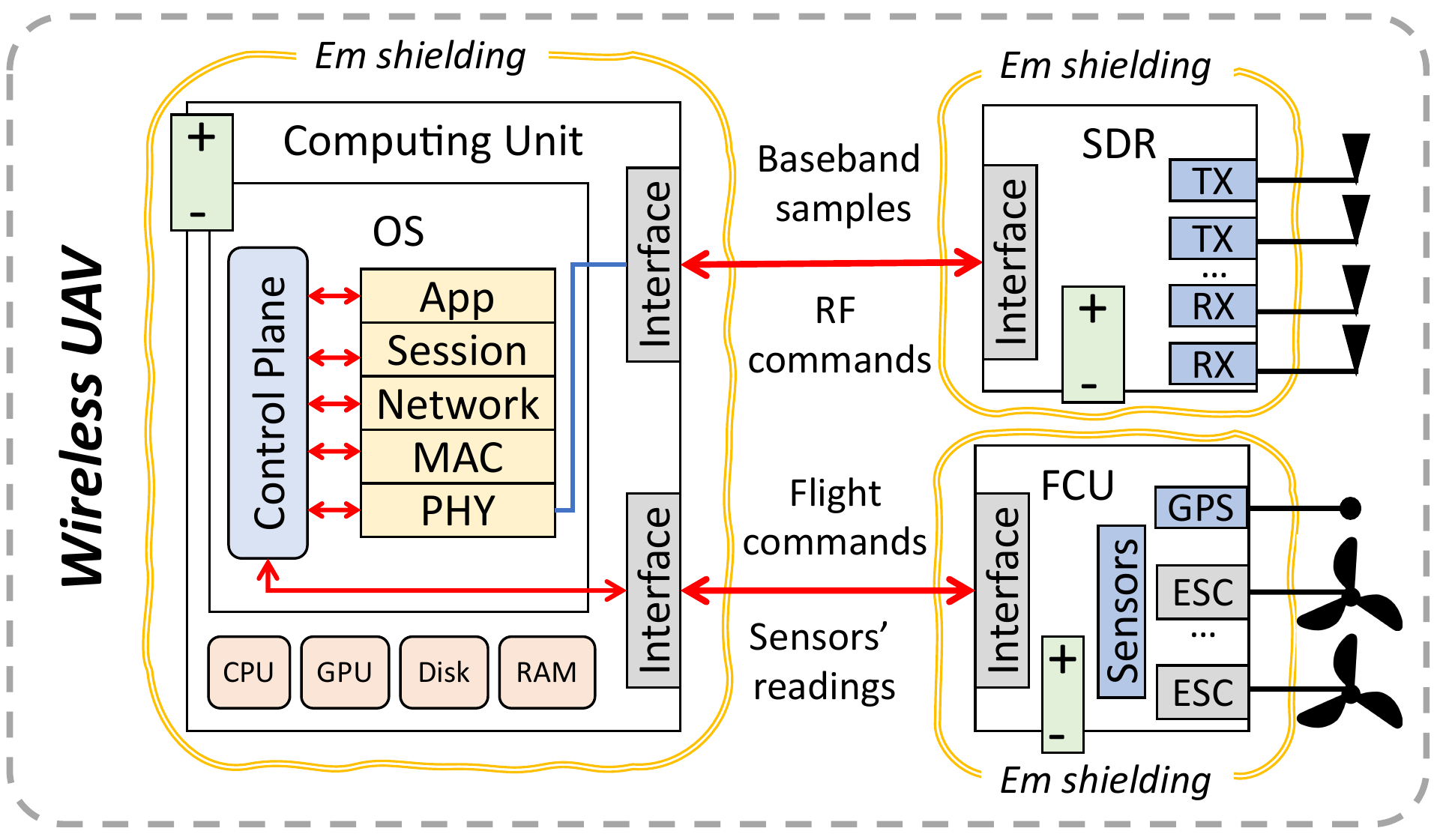} 
    \caption{Design blueprint for 6G wireless UAVs.}
    \label{fig:blueprint}
\end{figure}

An illustration of the design blueprint for future UAVs is given in Figure~\ref{fig:blueprint}.

\subsection{UAV design fundamentals} 
\label{sec:uav_fund}

To facilitate the design process of wireless UAVs, in here we introduce some notation.
Let $P_T$ be the total average power consumption of a wireless UAV during a flight. $P_T$ is expressed as:
\begin{align}
    P_T = P_F + P_C + P_R + P_L, \label{eqn_power}
\end{align}
where $P_F$ is the average required flight power (the electrical power required to keep the UAV air-born), $P_C$ is the average required computation power (the power required from the UAV's on-board computers and micro-controllers), $P_R$ is the average required radio power (power required to function the radio front end), and $P_L$ is the average lost power. 
Accordingly, the total flight time of a UAV $t_f$ can be found as a function of the total energy capacity on-board the UAV $E_T$ with the equation:
\begin{align}
    t_f = \frac{E_T}{P_T} = \frac{E_T}{P_F + P_C + P_R + P_L}. \label{eqn_flight_time}
\end{align}
Let's define the All Up Weight (AUW) of the UAV as the total weight of the UAV at the time of flight. This includes the weight of the frame, battery, computing unit, radios, and any attached payloads to the airframe (expressed in [$\mathrm{kg}$]). 
For a rotor-craft UAV (a  heavier-than-air aircraft that flies thanks to the lift generated by one or more rotors\cite{FAA-14-1}), 
the AUW is a good approximation of the average thrust $\Bar{T}$ (expressed in Kilogram-force [$\mathrm{kgf}$]) needed from the motors to keep the UAV up in the air
\begin{align}
    AUW \equiv \Bar{T} \label{eqn_auw}.
\end{align}
From the average thrust, the average per-motor thrust ($\Bar{T}_{m}$) can be found from the equation:

\begin{align}
    \Bar{T}_{m} = \Bar{T} / n, \label{eqn_motor_thrust}
\end{align}
where $n$ is the number of motors on the UAV (e.g., $n=4$ for quadcopters, $n=6$ for hexacopter). Conversely, the total average thrust can be found from the inverse equation:

\begin{align}
    \Bar{T} = \Bar{T}_{m} * n \label{eqn_total_thrust}.
\end{align}
%
Finally, the design of a wireless UAV must ensure that the payload can successfully be lifted with the available motors' thrust. 
A condition to ensure safe flying operations of the UAV is that the thrust generated by the motors at $100\%$ throttle must be greater than twice the AUW \cite{biczyski2020multirotor}.
To guarantee that a $2:1$ thrust-to-weight ratio, the design must ensure that the AUW is always below the thrust output of the motors at $60\%$ throttle or at an ESC's Pulse Width Modulation (PWM) of 1600, denoted by $T_m^{1600}$:
\begin{align}
    max(AUW) \leq n * T_m^{1600}.
\end{align}
The thrust generated by an individual motor (and its mounted propeller) is a function of the electrical power supplied to the motor itself. 
The thrust force versus power supply analysis can be performed via Computational Fluid Dynamics (CFD) on the propeller's airflow moved by the motor's torque at different speeds, using the motor performance data provided by the manufacturer, or acquired via the use of thrust stands such as \cite{rcbench_web}. 
From the motor's performance data, it is possible to determine the discrete values of the motor thrust $T_m(P_m)$ as a function of electrical power. Conversely, the discrete values of the motor's electrical power, $P_m(T_m)$, can be determined as a function of thrust. 
Finally, the total average flight power $P_F$ can be determined by adding the individual motor's powers. Substituting in Eqs. \ref{eqn_motor_thrust} and \ref{eqn_auw} we obtain:
\begin{align}
    P_F \approx n*P_m(T_m) = n*P_m(T/n) = n*P_m(AUW / n).
\end{align}
From this and Eq. \ref{eqn_flight_time}, the total flight time $t_f$ can be approximated as (without loss of generality we assume $P_L$ negligible):
\begin{align}
    t_f \approx \frac{E_T}{n*P_m(AUW / n) + P_C + P_R}.
\end{align}
For what concerns the onboard energy, the most common power source used on modern UAVs are Lithium Polymer (LiPo) or Lithium Ion (LIon) batteries. LiPo and LIon batteries cannot be discharged for their entire energy capacity due to their internal chemistry. Typically, only approximately $80\%-90\%$ of a LiPo battery capacity can be used without damaging the cells. Thus, only approximately $80\%$ of the battery capacity on-board a UAV is usable. The equation for the flight time of a multi-rotor UAV can then be further expressed as: 
\begin{align}
    t_f \approx \frac{0.8 * Batt_{Wh}}{n*P_m(AUW / n) + P_C + P_R}, \label{eqn_flight_curr}
\end{align}
where $Batt_{Wh}$ is the battery energy capacity in Watts hour. 

We will use the relationship between the UAV flight time, the battery capacity, and power consumption expressed in Eq. \ref{eqn_flight_curr} in the design process of a new wireless UAV prototype in the next section.


\subsection{UAV prototype}



The design process of prototyping a wireless UAV is iterative. 
As seen in the previous section, the choice of motors, propellers, and battery affect the AUW of the UAV, the thrust-current efficiency, and the total energy capacity on board. 
Indeed, increasing the battery size will increase the overall capacity on-board the UAV, but also will increase the AUW, causing the hovering throttle value to increase. 
In addition, unless the motors, propellers, and batteries are being directly manufactured, commercial equipment must be used, which limits the range of hardware to a few off-the-shelf models only.
\begin{figure}[b]
    \centering
    \includegraphics[width=\columnwidth]{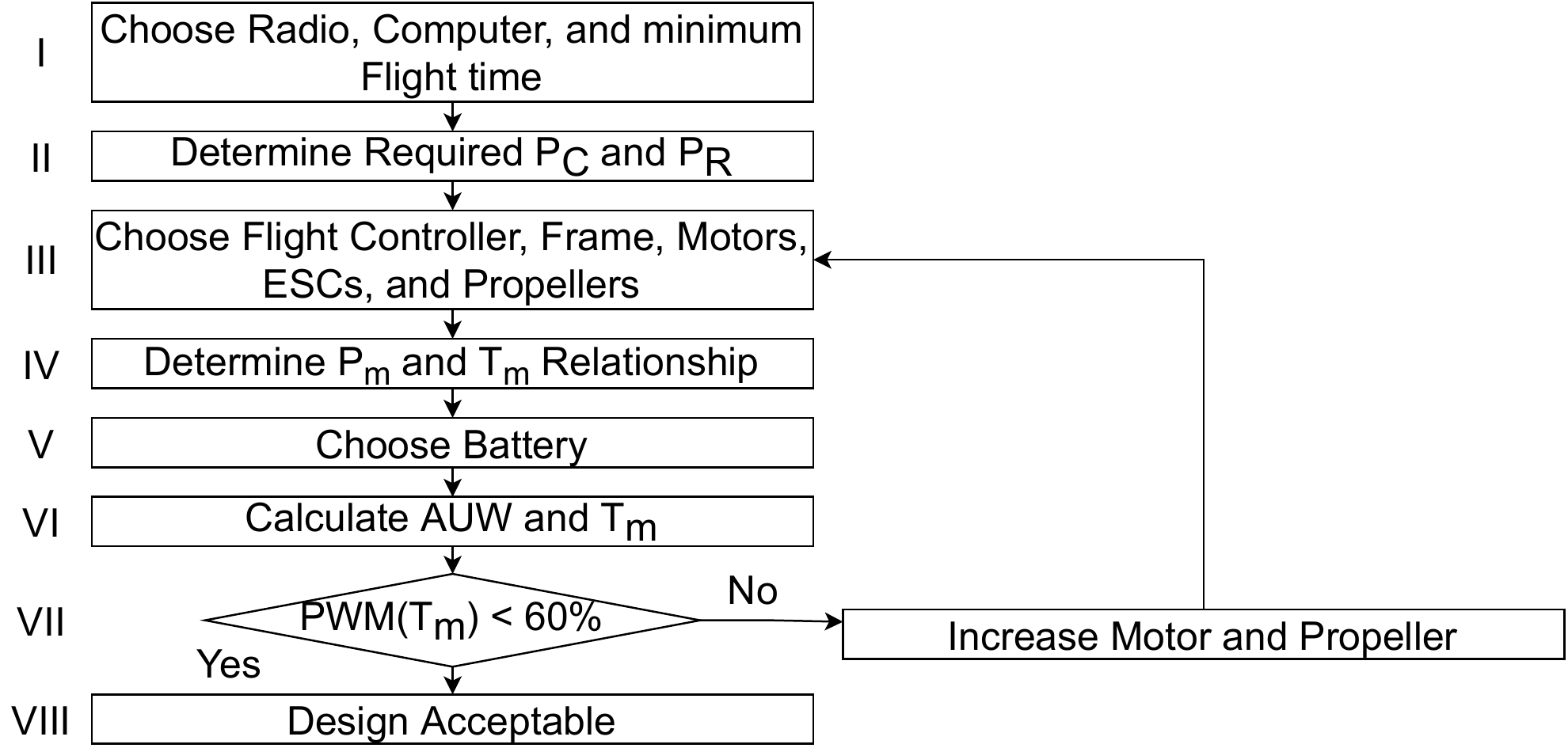} 
    \caption{Wireless UAV design process.}
    \label{fig:uav_design_proc}
\end{figure}
We use the iterative design process reported in Figure~\ref{fig:uav_design_proc} as a guideline to design a new wireless UAV prototype.
In there, we approximate the motor's PWM, the UAV's flight time, and AUW as  described in Section \ref{sec:uav_fund}. 

\begin{enumerate}[wide, labelwidth=!, labelindent=0pt, label=\textbf{(\Roman*)}]
    \item \label{lab:payload} \textit{Choose Radio, Computing Unit, and minimum Flight Time}:
    We have highlighted the importance of programmable hardware such as SDRs in future wireless UAVs for 6G spectrum research. 
    UAVs equipped with SDRs could implement different solutions, from aerial BSs\cite{BertizzoloMmnets19}, to aerial UEs\cite{bertizzolo2021streaming, BertizzoloHotmobile20}, to ad hoc UAV-to-UAV communications \cite{BertizzoloInfocom20SwarmControl}.
    To accomplish this goal, we chose NI's USRP B210 as a radio front end. USRP B210 ($\kg{0.35}$) is a fully-programmable lightweight software radio module featuring $\MHz{70}$ – $\GHz{6}$ carrier frequency range, $\MHz{56}$ of real-time bandwidth, 2 TX and 2 RX chains with MIMO capabilities. 
    \newline
    To operate USRP B210 and the wide range of supported software-based wireless applications (e.g., Open Air Interface, srsLTE, GNU Radio), we select the Intel NUC NUC7i7DN ($\kg{0.47}$) \cite{nuc}. 
    The Intel NUC is a commercial Mini PC, whose compact dimensions and good computational capabilities (Intel Core i7 CPU with $32\:\mathrm{GB}$ RAM) make it particularly suitable even to be carried on board of an UAV. 
    Finally, we specify a desired flight time of $45$ minutes.
    
    \item \label{lab:pcandpr} \textit{Determine Required $P_C$ and $P_R$}:
        From the products' data sheets, the maximum power requirements for the computer ($P_C$) and radio ($P_R)$ are 65W and 18W respectively \cite{nuc, b200}.
    
    \item \label{lab:motor} \textit{Choose Flight Controller, UAV frame, Motors, ESCs and Propellers}:
        The choice of the UAV frame, UAV's motors and propellers is subject to the availability of off-the-shelf products. 
        We chose a combination that can fit the chosen computing and radio hardware onboard yet containing the UAV's form factor and cost. 
        This choice can be iterated if necessary, as we will see later. The motor must be chosen first. 
        From the motor manufacturer's specification or suggestion, we can choose the battery voltage, propeller size, and ESC voltage.
        For our design, we first considered the T-Motor MN501-S KV240 motor ($\kg{0.17}$ each) with T-Motor P20*6.5 Propellers ($\kg{0.044}$ each) operating with an 8 cell ($\V{33.6}$) LiPo using TMotor Flame $\A{60}$ ESCs ($\kg{0.074}$ each) \cite{tmotor_mn501, tmotor_g20x6.5, tmotor_flame}. 
        As a UAV frame, we chose the IFlight IXC15 frame ($\kg{0.643}$) as it easily accommodate the motors and propellers, yet leaving enough room to mount radio and computing unit on board.
        We finally select a small form-factor Pixhawk Mini (45.2g) as FCU \cite{holy_pix_mini}.
        \begin{figure}[t]
            \centering
            \includegraphics[width=0.8\columnwidth]{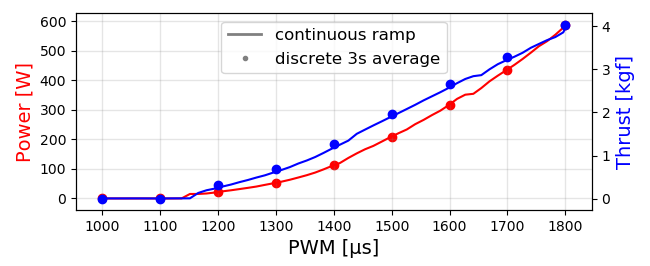}
            \caption{T-Motor MN501-S power and thrust analysis.}
            \label{fig:mn501}
        \end{figure} 
    \item \label{lab:PF} \textit{Determine the $P_m$ and $T_m$ Relationship}:
        To determine the non-linear relationship between the motors' input power and the motors' generated thrust, we preformed a static motor thrust analysis using an RCBenchmark Series 1580 Thrust Stand \cite{rcbench_web}. Specifically, we measure the motors' input power and the motors' output thrust as a function of PWM.
        Characterizing this relationship is necessary to calculate the battery capacity and weight, as we will see later.
        We perform a series of discrete measurements and report the interpolated power and thrust results in Figure~\ref{fig:mn501}. 
    \item \label{lab:battery} \textit{Choose the Battery}:
         From Eq. \ref{eqn_flight_curr} we can express the required battery capacity as a function of the motor input power as:
            \begin{align}
                Batt_{Wh} \geq&\frac{\left( n*P_m(PWM) +P_C + P_R \right)* t_f}{0.8 }\nonumber \\
                \geq&\frac{[4*P_m(PWM)+ \W{65} + \W{18}]}{0.8}*\frac{45\:min}{60\:min}.\label{eqn:batt_capacity}
            \end{align}
        Similarly, from Eqs. \ref{eqn_auw} and \ref{eqn_total_thrust}, we can express the required battery weight as a function of the thrust generated by the motors as:
        \begin{align}
               Batt_{kg} \leq & \Bar{T} - \text{AUW}_\text{other} = n*T_m(PWM) - \text{AUW}_\text{other}\nonumber\\
               \leq& n*T_m(PWM) -[ 4*(170g+44g+74g)+643g \nonumber\\
               &+45.2g+470g+350g]\nonumber\\
               \leq&4*T_m(PWM)-\kg{2.66}. \label{eqn:batt_weight}
            \end{align}
        By employing the relationship between motors' power and motors' generated thrust derived in (IV), we can calculate $Batt_{Wh}$ and $Batt_{kg}$ both as function of PWM and plot them against each other. We report this relationship in Figure~\ref{fig:battery_capacity}.

        \begin{figure}[ht]
            \centering             \includegraphics[width=0.8\columnwidth]{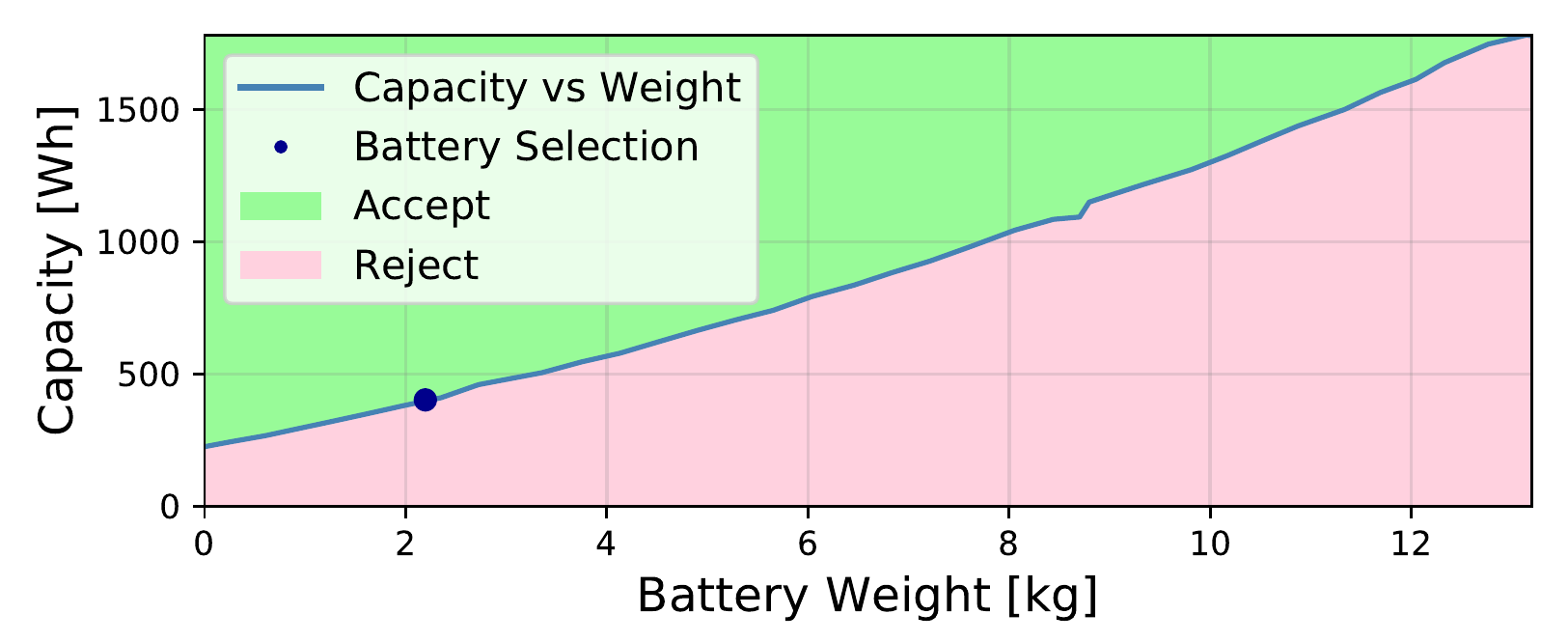} 
            \caption{Battery capacity vs.\ battery weight.}
            \label{fig:battery_capacity}
        \end{figure}      
        
        To satisfy Eqs. \ref{eqn:batt_capacity} and \ref{eqn:batt_weight}, any battery above the curve in  Figure~\ref{fig:battery_capacity} is a good choice. For our design, we chose to use 4 8S 6Ah LiPo batteries ($\kg{0.549}$ each) \cite{8s_bat} reported as a dot in Figure~\ref{fig:battery_capacity}. 
        From the power and thrust characterization data shown in Figure~\ref{fig:mn501}, we report the required Battery Capacity as a function of Battery weight in Figure~\ref{fig:battery_capacity}.
    \item \label{lab:auw_and_tf} \textit{Calculate AUW and $T_m$}:
        Following the battery selection, all the wireless UAV components have been selected. Thus, the theoretical AUW and $T_m$ can be calculated as:
        \begin{align*}
            AUW &\approx (\kg{170} +\kg{0.044}+\kg{0.074})*4 + \kg{0.643} \\
            &+ \kg{0.045} + \kg{2.199} + \kg{0.47} + \kg{0.350}= \kg{4.856}\\
            T_m &\approx AUW/4 = \kgf{1.214}.
        \end{align*}
    \item \label{lab:pwmCheck} \textit{Check Maximum PWM}:
     From the motor analysis shown in Figure~\ref{fig:mn501} we find that the individual motors' PWM required to lift the AUW is 
     $\text{PWM}(T_m=\kgf{1.214})\approx \us{1260} $. 
     Since the found PWM value is less than $1600 \mu s$, the design process outlined in Fig.\ref{fig:uav_design_proc} is satisfied and our prototype is completed.
\end{enumerate}

\begin{figure}[t]
    \centering
    \includegraphics[width=0.9\columnwidth]{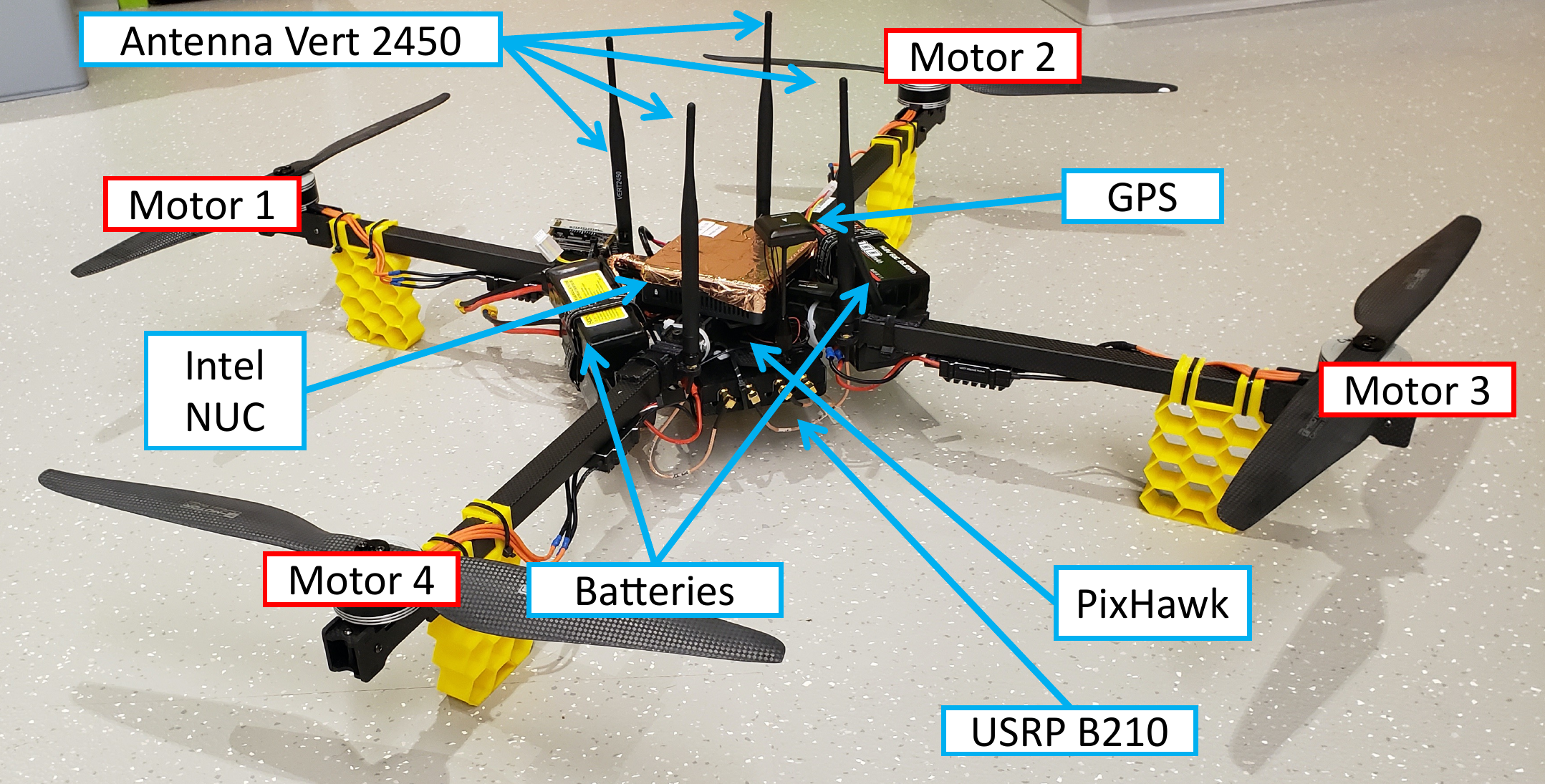} 
    \caption{The Monarch wireless UAV.}
    \label{fig:monarch}
\end{figure}

Last, we select Ardupilot as flight controller firmware (FCF), we equip the wireless UAV with 3D-printed landing gear and antenna mounts, use 4 VERT 2450 antennas for over-the-air communication and shield the wireless UAV components with copper foil to limit the electromagnetic noise that could impair RF and flight operations. The prototyped wireless UAV, termed \textit{Monarch}, is shown in Figure~\ref{fig:monarch}.
Its design, landing gear CAD models, mounting instructions, and components publicly available~\cite{wines-repo}.

\section{Experimental Performance Analysis}

We benchmarked the flight time performance of Monarch as well as its wireless capabilities and electromagnetic shielding though a series of real-world flight experiments.

\begin{figure}[t]
  \centering
  \subcaptionbox{
  Power and flight time vs. AUW.
  \label{fig:theory_flight_time}
  }
  {
    \includegraphics[width=.65\columnwidth,
    height=0.28\columnwidth
    ]
    {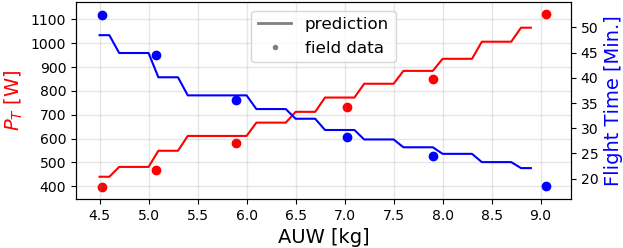}
  }
  \hfill
  \subcaptionbox{
    Flight autonomy benchmarking. 
    \label{fig:ft_comparison}
  }
  {
    \includegraphics[width=.3\columnwidth, 
    height=0.28\columnwidth
    ]
    {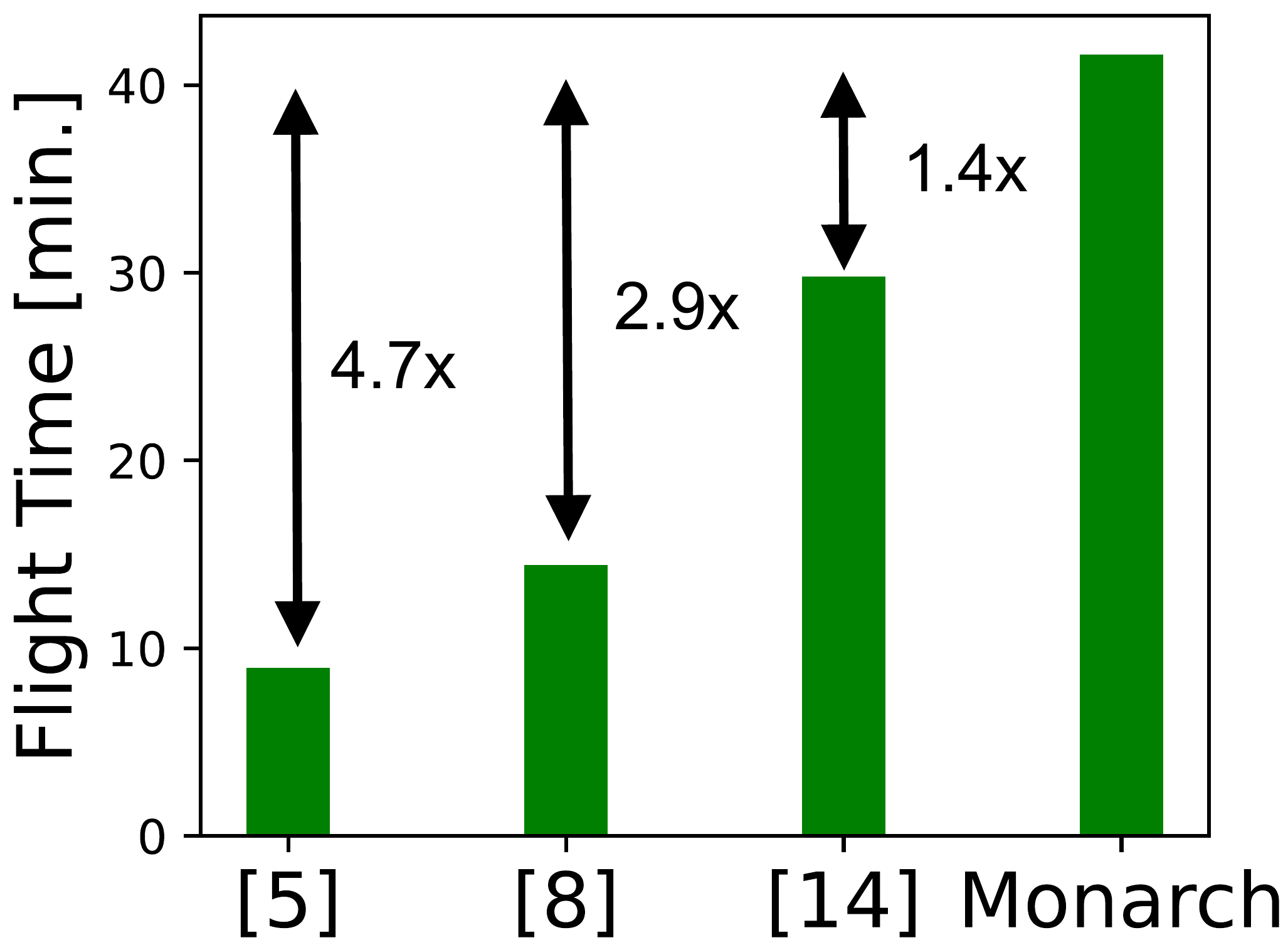}
  }
  \caption{Monarch flight autonomy and power performance.} 
\end{figure}

From the thrust and power analysis in Figure~\ref{fig:mn501}, we calculated the flight time prediction for different AUWs. 
We report this analysis in Figure~\ref{fig:theory_flight_time}. 
To support our calculations, we experimentally verified the UAV's flight time relative to different payload weights through real-world flight experiments.
To do this, we manually added variable mass payload to the UAV and autonomously flight the UAV in a set square pattern until the battery voltage under load went below $3.6V$ per LiPo cell. 
The field testing results are consistent with our analysis and are also reported in Figure~\ref{fig:theory_flight_time}. We recorded a flight time of 44.43 minutes for AUW of $\kg{5.08}$ with an onboard NUC and B210. 
Compared to off-the-shelf UAV models equipped with wireless modules, Monarch provides superior flight time.  Our constructed design performance are consistent with prediction, thus we validate our design and our design methodology. 

\begin{figure}[b]
    \centering
    \includegraphics[width=0.8\columnwidth]{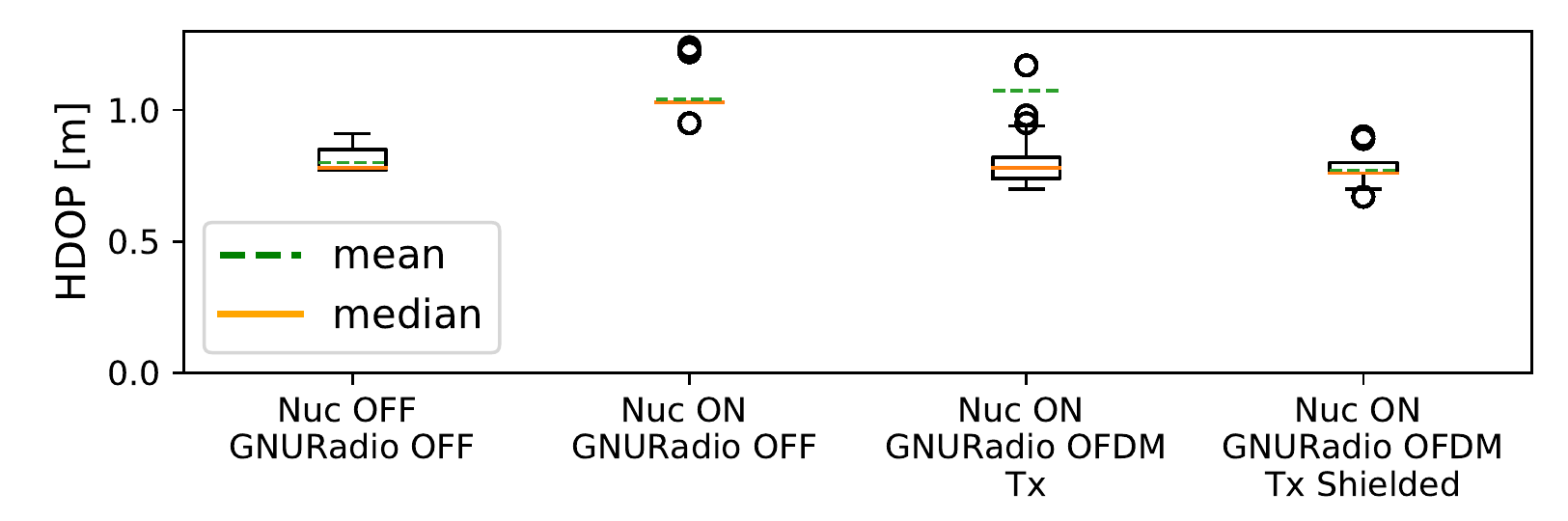}
    \caption{GPS Horizontal Dilution of Precision (HDOP) for different wireless applications.}
    \label{fig:interference}
\end{figure} 

We underline the importance of prototyping wireless UAVs adopting the presented bottom-up approach by benchmarking Monarch's flight autonomy with relevant work in literature.  
We compare Monarch with similar off-the-shelf UAV models equipped with wireless modules in Fig. \ref{fig:ft_comparison}.
When compared to \cite{BertizzoloInfocom20SwarmControl}, \cite{bertizzolo2021streaming}, and \cite{ferranti2020skycell}, Monarch reports flight autonomy gains of $4.7$x $2.9$x, and $1.4$x, respectively. 

Last, we performed an experimental validation of the electromagnetic noise figure on board. To do so, we used the GPS Horizontal Dilution of Precision (HDOP) as an indicator of the electromagnetic noise figure to which the on-board sensors are subject to.
Specifically, we have measured the HDOP over 3-minute long flight experiments with different wireless applications running onboard.
\begin{enumerate*}[label=(\roman*)]
    \item \textit{NUC off, Radio off}: all electronics on board is powered off. This case represents ideal conditions.
    \item \textit{NUC on, Radio off}: The computing unit (NUC) is turned on and runs the OS. No electromagnetic isolation is applied here.
    \item \textit{NUC on, Radio on (Gnuradio OFDM TX)}: The computing unit (NUC) and the SDR (USRP B210) are on and run an OFDM transmitter at $\GHz{2.4}$ with $\MHz{1}$ of bandwidth. No electromagnetic isolation is applied here.
    \item \textit{NUC on, Radio on (Gnuradio OFDM TX) with shielding}: The computing unit (NUC) and the SDR (USRP B210) are on and run an OFDM transmitter at $\GHz{2.4}$ with $\MHz{1}$ of bandwidth. We do apply electromagnetic isolation here.
\end{enumerate*}
The results are reported in Figure~\ref{fig:interference}.
Without isolation, the computing unit and radio application rise the average HDOP from $\m{0.8}$ to $\m{1.04}$ and $\m{1.07}$, respectively, with outliers as high as $\m{3}$. 
On the other hand, the applied copper shielding provides electromagnetic isolation with respect to the on-board computing and wireless application and features an even lower mean HDOP than when all electronics are powered off ($\m{0.77}$). 


\section{Open Challenges}

As we are still in the early stages of aerial wireless communications and wireless UAVs development, we conclude this article with a few major challenges to keep in mind for the design and development of future wireless UAVs.

\begin{itemize}[wide, labelwidth=!, labelindent=0pt]
    \item \textit{Small form-factor UAVs and long-range wireless communications}:
    Long-range communications with in-orbit 6G satellites or users on the ground might require significant transmission power and large antenna gains. The latter can be achieved with large  antenna modules or high order antenna arrays. Large antenna modules (e.g., log-periodic, yagi, or horn antennas) are bulky and might take significant room on board. Despite these modules are passive and do not require significant power, they are generally heavy and can reduce the flight time when carried onboard. 
    On the other hand, high-order antenna arrays can be used with analog and digital beamforming techniques to `convey' the signal energy toward a single direction and extend the communication range. However, these modules require a clear mounting surface and consume a significant amount of power. 
    Additionally, both solutions have a limited coverage angle, which requires intelligent steering mechanisms or employing multiple modules to extend coverage.
    Functional antenna design, such as curved and frame-embedded antennas, similar to the ones in modern smartphones and tablets, can help equip wireless UAVs with high-gain antenna modules and address some of these issues.

    \item \textit{Programmable radio front-end and software stack powering}: 
    As hardware programmability is a must-have feature for future non-terrestrial networks, efficient powering of high-performance programmable hardware is still an open question. 
    High-bandwidth and high-power programmable radio front ends are still unsuitable to be mounted on board of a UAV. Similarly, 
    high-performance computing units necessary to process the baseband processing at ultra-wide bands are large-form-factor server-like computing units, unfit to be carried on board.
    Hardware and software advances in power-efficient protocol suites implementations as well as power-efficient computing units and low-power SDR boards will ease the portability of programmable radio front ends on small UAVs.

    \item \textit{UAV frame blockage and shadowing}: 
    One key design choice in the prototyping of a wireless UAV is the placement of RF antennas on board. To maintain a small-form factor and stable flight operations, these are generally placed above or below the main frame. 
    In close proximity with the antennas, the UAV frame can `shield' the signal coming in and out the RF antennas and weaken the UAV communication capabilities. This effect is known as `frame shadowing' and depends on the frame's material, size, and the wireless transmission direction. 
    This problem is exacerbated when implementing wireless communications with UAVs hovering at different altitudes. 
    Employing multiple antennas together with intelligent antenna selection or antenna steering mechanisms can help alleviate this problem.

    \item \textit{In-motion communications}:
    The popularity of wireless UAVs can be found in their ability to combine 3D mobility with wireless communication.
    While many works have taken advantage of the combination of these two aspects, less effort has been put in studying the problematic interaction of aerial mobility and wireless communications.
    Involuntary micro-mobility, involuntary fluctuation, and communications at different altitudes, can degrade or even disrupt a wireless link. 
    Even more problematic, the very principle at the base of UAV's mobility---tilting---can inadvertently change the antenna polarization, significantly change the boresight of a directional antenna module, or impair the effectiveness of digital and analog beamforming techniques.
    Thus, developing intelligent in-flight transmission mechanism is fundamental to support in-motion wireless communications for UAVs. 
    Several approaches are possible. These can be hardware implementations (mobile antennas, antenna gimbals, phased array compensations, antenna selection) or software implementations (e.g., motion-adaptive beamforming). 
    Both cases can greatly benefit from accessing the FCU sensors' readings such as the accelerometer, the Inertial Measurement Units (IMU), and the Global Positioning System (GPS).
\end{itemize}

\section{Conclusions}

This article introduces a formal definition of a wireless UAV.
We revise key design choices for future 6G wireless UAVs and present Monarch, a new wireless UAV prototype for 6G spectrum research. 
The Monarch design is publicly available~\cite{wines-repo}.
Our work concludes indicating a few open challenges in the domain of UAV-based wireless communications.

\balance
\bibliographystyle{abbrv}
\bibliography{biblio}

\end{document}